\documentclass{jnmp}

\usepackage{amsmath}

\JNMPnumberwithin{equation}{section}

\begin{document}

\renewcommand{\evenhead}{V~E Vekslerchik}
\renewcommand{\oddhead}{Functional Representation of the Ablowitz--Ladik Hierarchy. II}

\thispagestyle{empty}
\FirstPageHead{9}{2}{2002}{\pageref{Vekslerchik-firstpage}--\pageref{Vekslerchik-lastpage}}{Article}

\copyrightnote{2002}{V~E~Vekslerchik}

\Name{Functional Representation \\
  of the Ablowitz--Ladik  Hierarchy. II}\label{Vekslerchik-firstpage}

  \Author{V E VEKSLERCHIK}

\Address{Departamento de Matem\'aticas, E. T. S. I. Industriales, \\
  Universidad de Castilla-La Mancha, \\
  Avenida de Camilo Jos\'e Cela, 3, 13071 Ciudad Real, Spain\\[5pt]
On leave from the Institute for Radiophysics and Electronics, Kharkov, Ukraine\\[5pt]
Regular Associate of the   Abdus Salam ICTP,  Trieste, Italy}

\Date{Received July 27, 2001;
  Revised October 31, 2001;
  Accepted December 17, 2001}

\begin{abstract}
\noindent
In this paper we continue studies of the functional
representation of the Ablowitz--Ladik hierarchy (ALH). Using formal
series solutions of the zero-curvature condition we rederive the
functional equations for the tau-functions of the ALH and obtain
some new equations which provide more straightforward description
of the ALH and which were absent in our previous paper. These
results are used to establish relations between the ALH and the
discrete-time nonlinear Schr\"odinger equations, to deduce the
superposition formulae (Fay's identities) for the tau-functions of
the hierarchy and to obtain some new results related to the Lax
representation of the ALH and its conservation laws. Using the
previously found connections between the ALH and other integrable
systems we derive functional equations which are equivalent to the
AKNS, derivative nonlinear Schr\"odinger and Davey--Stewartson
hierarchies.
\end{abstract}

\section{Introduction.}

In \cite{1} we have presented some results on the
functional representation of the Ablowitz--Ladik hierarchy (ALH).
The ALH, which had been originally proposed as an infinite set of
the differential-difference equations (see \cite{AL1}), has been
written as a finite system of functional-difference equations. The
aim of this work is to continue these studies. However, to present
some formulae which were absent in~\cite{1} (such as, first of
all, the functional relations written in terms of the solutions
themselves, and not in terms of the corresponding
tau-functions) is not the only (and not even the main) goal of
this work. We wish to fill some gaps in the theory of
the ALH. The ALH, which was discovered by Ablowitz and Ladik a
quarter of century ago, is one of the first integrable discrete
$(1+1)$-dimensional systems. Nevertheless, maybe because of lack of
interesting practical applications, it has not received proper
attention from mathematicians and physicists. In the not so distant past
the `classical' inverse scattering transform (IST) of~\cite{AL1}
and Hirota's bilinear method, which were elaborated in the seventies,
were almost the only tools for investigating the ALH. Other
techniques developed during the last two decades (such as, e.g.,
formal series, Sato's approaches, discrete-time representation
etc) have not been adopted to this integrable system. (Only
recently a few works have appeared in which the ALH is considered from
more modern viewpoints, see, e.g.,~\cite{MEKL,Suris}.) The paper~\cite{1}
was not an exception. The method of~\cite{1} is the
method of the pioneering article~\cite{AL1}: the main equations
of~\cite{1} were obtained by analyzing the structure of the Jost
functions of the auxiliary scattering problem. The Jost functions,
by definition, crucially depend on the boundary conditions
imposed, while the form of the equations of the hierarchy, and
hence the resulting functional equations, do not. Here we
exploit another approach which is based on the local properties of
the ALH and will enable us to avoid restrictions caused by the
boundary conditions (the proof of main results in~\cite{1}, but
not the results themselves, was limited to the case of zero
boundary conditions). This also gives us the possibility to see
how the main objects of modern theory of integrability appear in
the case of the ALH.

We start with the standard zero-curvature (or the so-called
$U$--$V$) representation (ZCR) based on the $U$-matrix of the
work~\cite{AL1}, but instead of considering the polynomial, in
auxi\-liary parameter $\lambda$, of $V$-matrices $V_{n}^{j}$ (as
is in `classical' IST, \cite{AL1}) we deal with some formal matrix
series from which the former can be obtained by multiplication
plus projection scheme, $V_{n}^{j}(\lambda) = \pi_{\lambda}^{+}
\left[ \lambda^{-2j} \; \mathcal{V}_{n}(\lambda) \right]$ (the
definitions of $V_{n}^{j}(\lambda)$, $\mathcal{V}_{n}(\lambda)$,
$\pi_{\lambda}^{\pm}$ as well as basic facts related to the ZCR
one can find in Section~\ref{sec-zcr}). This approach is widely
used in the theory of the KdV, KP and other hierarchies but, to my
knowledge, has not been elaborated in the~case of the ALH. Then we
show how the operator $\lambda \, \mathrm{d} / \mathrm{d}\lambda
$, the importance of which has been discussed in the literature
(see, e.g.,~\cite{Newell}), acts on these series, which in a
natural way leads to the so-called Miwa's shifts, $z_{k} \to z_{k}
+ \varepsilon \zeta^{k}/k$~\cite{Miwa}. The Miwa's shifts are one
of the most characteristic attributes of the modern theory of
integrability, which in such or~another form are present in
considerable part of recent works devoted to the KP (together with
its modifications), KdV, 2D Toda, Davey--Stewartson and other
hierarchies. However, in the context of the ALH they, to my
knowledge, have not been discussed in the literature. After
deriving the functional equations which are equivalent to the ALH
(Section~\ref{sec-fe}) and rewriting them in terms of the
tau-functions of the ALH (Section~\ref{sec-tau}) we address the
question of the relations between these equations and the
discrete-time Ablowitz--Ladik equations~\cite{AL3,Suris}. Having
established the discrete-time character of the functional
representation of the ALH, i.e.\ considering the combined action
of the evolutionary ALH flows as the integrable mapping of the
work~\cite{Suris}, it is rather natural to address the problem of
superposition of these maps (i.e.\ superposition of the Miwa's
shifts). In Section~\ref{sec-fay} we present Fay's formulae for
the ALH which relate functions calculated at different values of
arguments, $z_{k} + \sum_{a} \varepsilon_{a} \zeta_{a}^{k}/k$, and
this in turn provides some interesting results related to the
question of the higher-order discrete-time Ablowitz--Ladik
equations~\cite{Suris2} (see Section~\ref{sec-hdf}).

Thus in this paper we explicitly construct the chain from the
scattering problem of Ablowitz and Ladik (written in 1975) through
formal series and Miwa's shifts to the discrete-time equations
written by Suris in 1997. However, it seems useful before we move
forward (say, to the Grassmannian description of the ALH, which,
again, is still to be developed) that we return to some
`classical' problems and demonstrate that the functional
representation possesses some practical value. The key moment is
that now we know the (formal) solution of the auxiliary linear
problems, which enables us to obtain some new expressions for the
results which have been formulated in terms of Jost functions and
the scattering matrices. To this end we discuss once more the
question of conservation laws. The form of the generating function
for the divergent-like conservation laws presented in
Section~\ref{sec-cl} seems to be new and rather convenient for practical
usage.

Lastly we demonstrate that some of the ideas and results
of this paper can be useful beyond the theory of the ALH. To do
this we give in Section~\ref{sec-examples} few examples of
the `embedding into the ALH' approach: starting from the ALH
functional equations we derive the functional representation
of some other integrable models, namely, the derivative nonlinear
Schr\"odinger (NLS), AKNS and Davey--Stewartson hierarchies.

\section{Zero curvature representation of the ALH} \label{sec-zcr}

The ALH is an infinite set of differential-difference equations,
which can be presented as the compatibility condition for the linear
system
\begin{gather}
\Psi_{n+1} = U_{n} \Psi_{n},
\label{zcr-sp}\\
\partial \Psi_{n} = V_{n} \Psi_{n},
\label{zcr-evol}
\end{gather}
where $\Psi_{n}$ is a $2$-column (or $2 \times 2$ matrix),
$U_{n}$ and  $V_{n}$ are $2 \times 2$ matrices with $U_{n}$ being given by
\begin{equation}
U_{n} = U_{n} (\lambda) =
\begin{pmatrix} \lambda & r_{n} \\ q_{n} & \lambda^{-1} \end{pmatrix}
\end{equation}
(here $\lambda$ is an auxiliary constant parameter). To provide
the self-consistency of the system~(\ref{zcr-sp}),~(\ref{zcr-evol})
the matrices $V_{n}$ have to satisfy the
following equation
\begin{equation}
\partial U_{n} = V_{n+1} U_{n} - U_{n} V_{n}. \label{ZCR}
\end{equation}

One can divide the ALH into two subsystems (subhierarchies): the
`positive' one (with the $V$-matrices being polynomials in
$\lambda^{-1}$) and the `negative' one (with the $V$-matrices being
polynomials in $\lambda$). Since these subhierarchies are rather similar,
we concentrate mostly on the `positive' subsystem which stems
from (\ref{zcr-sp}) and
\begin{equation}
\partial_{j} \Psi_{n} \equiv
{\partial \over \partial z_{j}} \Psi_{n} =
V_{n}^{j} \Psi_{n},
\qquad j=1,2, \dots,
\label{zcr-evol-pos}
\end{equation}
where the matrices $V_{n}^{j}$ are polynomials of the $(2j)$th order in
$\lambda^{-1}$ with the following structure:
\begin{equation}
V_{n}^{j} = \lambda^{-2} V_{n}^{j-1} +
\begin{pmatrix} \lambda^{-2} \alpha_{n}^{j} & \lambda^{-1} \beta_{n}^{j} \vspace{1mm}\\
          \lambda^{-1} \gamma_{n}^{j} & \delta_{n}^{j} \end{pmatrix},
\qquad j=2,3, \dots.
\label{v-j}
\end{equation}
Substituting (\ref{v-j}) in the zero-curvature representation (\ref{ZCR})
one can obtain the set of equations determining the quantities
$\alpha_{n}^{j}, \ldots , \delta_{n}^{j}$
\begin{gather}
\alpha_{n}^{j} = - \delta_{n}^{j-1},
  \label{zcr-a-j}
\\
\beta_{n}^{j} = \beta_{n-1}^{j-1}
  + r_{n-1} \left( \delta_{n-1}^{j-1} + \delta_{n}^{j-1} \right),
\label{zcr-b-j}
\\
\gamma_{n}^{j} = \gamma_{n+1}^{j-1}  + q_{n} \left( \delta_{n}^{j-1} + \delta_{n+1}^{j-1} \right),
\label{zcr-c-j}
\\
\delta_{n+1}^{j} - \delta_{n}^{j} =
  q_{n} \beta_{n}^{j} - r_{n} \gamma_{n+1}^{j}
\label{zcr-d-j}
\end{gather}
and present equations of the hierarchy as
\begin{equation}
\partial_{j} q_{n} =
  q_{n} \delta_{n+1}^{j} + \gamma_{n+1}^{j}
\label{dq-j}
\end{equation}
and
\begin{equation}
\partial_{j} r_{n} =
  - r_{n} \delta _{n}^{j} - \beta_{n}^{j}.
\label{dr-j}
\end{equation}
The $V$-matrices of the `negative' subhierarchy, $\bar V_{n}$ (in this
paper the overbar \textit{does not} mean the complex conjugation!)
possess the following structure:
\begin{equation}
\bar V_{n}^{j} = \lambda^{2} \bar V_{n}^{j-1} +
\begin{pmatrix} \bar\alpha_{n}^{j} & \lambda \bar\beta_{n}^{j} \vspace{1mm}\\
          \lambda \bar\gamma_{n}^{j} & \lambda^{2} \bar\delta_{n}^{j} \end{pmatrix}
\end{equation}
and the equations of the subhierarchy can be written as
\begin{equation}
\bar\partial_{j} q_{n} =
  - q_{n} \bar\alpha_{n}^{j} - \bar\gamma_{n}^{j}
\label{bdq-j}
\end{equation}
and
\begin{equation}
\bar\partial_{j} r_{n} =
  r_{n} \bar\alpha_{n+1}^{j} + \bar\beta_{n+1}^{j},
\label{bdr-j}
\end{equation}
where $\bar\partial_{j} = \partial / \partial \bar z_{j}$. All
equations (\ref{dq-j}), (\ref{bdq-j}) and (\ref{dr-j}),
(\ref{bdr-j}) are compatible.  Thus we consider the $q_{n}$ and
$r_{n}$ as functions of the infinite number of variables $z_{j}$
and $\bar z_{j}$ and write, e.g.,
$q_{n}(z, \bar z)$ bearing in mind
\begin{equation}
  q_{n}(z, \bar z) =
  q_{n}(z_{1}, \bar z_{1}, z_{2}, \bar z_{2}, \dots).
\end{equation}

Till now we were following the standard zero-curvature scheme
of~\cite{AL1} (see also, e.g.,~\cite{AS}), but hereafter, since our
purpose is to discuss the ALH as a whole, we deal not with the
quantities $\alpha_{n}^{j}, \ldots , \delta_{n}^{j}$ (which describe
the~$j$th flow) but with series defined by
\begin{equation}
\begin{pmatrix}
  a_{n}(\zeta) & b_{n}(\zeta) \\
  c_{n}(\zeta) & d_{n}(\zeta) \end{pmatrix} =
\sum_{j=1}^{\infty} \zeta^{j}
\begin{pmatrix}
  \alpha_{n}^{j} & \beta_{n}^{j} \\
  \gamma_{n}^{j} & \delta_{n}^{j}\end{pmatrix}
\label{abcd}
\end{equation}
which describe simultaneously all `positive' flows.  In these terms
equations (\ref{zcr-b-j})--(\ref{zcr-d-j}) have the forms
\begin{gather}
b_{n+1}(\zeta) =
  \zeta b_{n}(\zeta) +
  \zeta r_{n} \left[ d_{n+1}(\zeta) + d_{n}(\zeta) - i \right],
\label{zcr-b}
\\
c_{n}(\zeta) =
  \zeta c_{n+1}(\zeta) +
  \zeta q_{n} \left[ d_{n+1}(\zeta) + d_{n}(\zeta) - i \right],
\label{zcr-c}
\\
d_{n+1}(\zeta)- d_{n}(\zeta) =
  q_{n} b_{n}(\zeta) - r_{n} c_{n+1}(\zeta)
\label{zcr-d}
\end{gather}
while equations (\ref{dq-j}) and (\ref{dr-j}) can be written as
\begin{equation}
\partial(\zeta) q_{n} =
  q_{n} d_{n+1}(\zeta) + c_{n+1}(\zeta)
\end{equation}
and
\begin{equation}
\partial(\zeta) r_{n} =
   - r_{n} d_{n}(\zeta) - b_{n}(\zeta),
\end{equation}
where
\begin{equation}
  \partial(\zeta) =
  \sum\limits_{j=1}^{\infty} \zeta^{j} \partial_{j}.
\end{equation}

It follows from (\ref{zcr-b})--(\ref{zcr-d}) that the series
$b_{n}(\zeta)$, $c_{n}(\zeta)$ and $d_{n}(\zeta)$ are not independent.
They are related by
\begin{equation}
  b_{n}(\zeta) c_{n}(\zeta) +
  \zeta d_{n}(\zeta) \left[ d_{n}(\zeta) - i \right] =
  \mathrm{constant}.
\label{bcd-constant}
\end{equation}
Using the invariance of the system (\ref{zcr-b})--(\ref{zcr-d}) with
respect to the transformations
\begin{equation}
b_{n} \to (1 + 2f) b_{n},
\qquad
c_{n} \to (1 + 2f) c_{n},
\qquad
d_{n} \to (1 + 2f) d_{n} - if
\end{equation}
(with arbitrary $f$) we can eliminate the constant in the
right-hand side of (\ref{bcd-constant}) and rewrite it as
\begin{equation}
  b_{n}(\zeta) c_{n}(\zeta) +
  \zeta d_{n}(\zeta) \left[ d_{n}(\zeta) - i \right] = 0.
\label{bcd}
\end{equation}
This choice corresponds to the situation when we define
$a_{n}^{0} = b_{n}^{0} = c_{n}^{0} = 0$,
$d_{n}^{0} = - i$
and do not introduce additional constants when we solve (\ref{zcr-d-j}) for
all $j$'s, which leads to the, so to say, simplest form of the ALH
equations (considered in~\cite{1}). Other choices lead to replacement of
the ALH flows with their linear combinations.

{\samepage Another consequence of (\ref{zcr-b})--(\ref{zcr-d}) is
that matrices $\mathcal{V}_{n}$ defined by
\begin{equation}
\mathcal V_{n}(\lambda) =
\begin{pmatrix}
   - d_{n}\left(\lambda^{2}\right)            &
   \lambda^{-1} b_{n}\left(\lambda^{2}\right) \vspace{1mm}\\
   \lambda^{-1} c_{n}\left(\lambda^{2}\right)  &
   d_{n}\left(\lambda^{2}\right) - i\end{pmatrix}
\label{calv-def}
\end{equation}
satisfy the equation
\begin{equation}
  \mathcal{V}_{n+1}(\lambda) U_{n}(\lambda) -
  U_{n}(\lambda) \mathcal{V}_{n}(\lambda)  = 0,
\label{zcr-stat}
\end{equation}
which is the stationary zero-curvature equation (\ref{ZCR}). Now one can
express the matrices $V_{n}^{j}$ describing the ALH flows as
\begin{equation}
V_{n}^{j}(\lambda) =
\pi_{\lambda}^{+}
\left[ \lambda^{-2j} \; \mathcal{V}_{n}(\lambda) \right],
\label{pi-plus}
\end{equation}
where $\pi_{\lambda}^{+}$ is a projection operator defined by
\begin{equation}
\pi_{\lambda}^{+}
\begin{pmatrix}
  \sum\limits_{j=-\infty}^{\infty} a_{j} \; \lambda^{2j} &
  \sum\limits_{j=-\infty}^{\infty} b_{j} \; \lambda^{2j+1} \vspace{1mm}\\
  \sum\limits_{j=-\infty}^{\infty} c_{j} \; \lambda^{2j+1} &
  \sum\limits_{j=-\infty}^{\infty} d_{j} \; \lambda^{2j}
\end{pmatrix}
=
\begin{pmatrix}
  \sum\limits_{j=-\infty}^{-1} a_{j} \; \lambda^{2j} &
  \sum\limits_{j=-\infty}^{-1} b_{j} \; \lambda^{2j+1} \vspace{1mm}\\
  \sum\limits_{j=-\infty}^{-1} c_{j} \; \lambda^{2j+1} &
  \sum\limits_{j=-\infty}^{0}  d_{j} \; \lambda^{2j}
\end{pmatrix}.
\end{equation}
Thus we have a usual situation for the case of `integrable
mathematics'. The space of the matrices involved in our hierarchy,
$G_{\lambda}$, can be in a natural way decomposed in two subspaces
$G_{\lambda} = \pi_{\lambda}^{+} G_{\lambda} \oplus
\pi_{\lambda}^{-}G_{\lambda}$ where $\pi_{\lambda}^{-}$ is given
by
\begin{equation}
\pi_{\lambda}^{-}
\begin{pmatrix}
  \sum\limits_{j=-\infty}^{\infty} a_{j} \; \lambda^{2j} &
  \sum\limits_{j=-\infty}^{\infty} b_{j} \; \lambda^{2j+1} \vspace{1mm}\\
  \sum\limits_{j=-\infty}^{\infty} c_{j} \; \lambda^{2j+1} &
  \sum\limits_{j=-\infty}^{\infty} d_{j} \; \lambda^{2j}
\end{pmatrix}
=
\begin{pmatrix}
  \sum\limits_{j=0}^{\infty} a_{j} \; \lambda^{2j} &
  \sum\limits_{j=0}^{\infty} b_{j} \; \lambda^{2j+1} \vspace{1mm}\\
  \sum\limits_{j=0}^{\infty} c_{j} \; \lambda^{2j+1} &
  \sum\limits_{j=1}^{\infty}  d_{j} \; \lambda^{2j}
\end{pmatrix},
\end{equation}
and all flows of the hierarchy can be constructed by projection
onto one of them.}

The matrix $\mathcal{V}_{n}(\lambda)$ plays a key role in the
considerations below since it can be used to construct
a (formal) solution of the linear system (\ref{zcr-sp}) and
(\ref{zcr-evol-pos}) (note that equation~(\ref{zcr-stat}) is
closely related to problem (\ref{zcr-sp})).

\section{Derivation of the functional equations} \label{sec-fe}

The aim of this section is to obtain the functional equations
which are equivalent to the ALH. As we mentioned in the
introduction, in this paper we use an approach different from that
of the work~\cite{1} and now we want to expose some differential
properties of the series~(\ref{abcd}) which are `hidden' in
the recurrence relation~(\ref{v-j}).

Using (\ref{dq-j}), (\ref{dr-j}) together with (\ref{zcr-b-j}),
(\ref{zcr-c-j}) one can obtain the following relations,
\begin{gather}
  \partial_{k}  \, \beta^{j}_{n} - \partial_{k+1}  \, \beta^{j-1}_{n}
  =
 \beta^{j}_{n} \, \delta^{k}_{n} +
  \beta^{j-1}_{n} \, \delta^{k+1}_{n} -
  2\delta^{j-1}_{n} \, \beta^{k+1}_{n},
\\
  \partial_{k} \, \gamma^{j}_{n} - \partial_{k+1} \, \gamma^{j-1}_{n}
  =
  - \gamma^{j}_{n} \, \delta^{k}_{n} -
  \gamma^{j-1}_{n} \, \delta^{k+1}_{n} +
  2\delta^{j-1}_{n} \, \gamma^{k+1}_{n}
\end{gather}
which, after replacing $j$, $k$ with $j+l$, $k-l$ and summing over $l$,
become
\begin{gather}
  \partial_{k} \, \beta^{j}_{n} =
  i \beta^{j+k}_{n} -
  \beta^{j}_{n} \, \delta^{k}_{n} +
  2 \delta^{j}_{n} \, \beta^{k}_{n} +
  2\sum_{l=1}^{j} \left(
    \beta^{l}_{n} \, \delta^{j+k-l}_{n} -
    \delta^{l}_{n} \, \beta^{j+k-p}_{n}
  \right),
\label{dbeta}
\\
  \partial_{k} \, \gamma^{j}_{n} =
  - i \gamma^{j+k}_{n} +
  \gamma^{j}_{n} \, \delta^{k}_{n} -
  2 \delta^{j}_{n} \, \gamma^{k}_{n} +
  2\sum_{l=1}^{j} \left(
    \delta^{l}_{n} \, \gamma^{j+k-l}_{n} -
    \gamma^{l}_{n} \, \delta^{j+k-l}_{n}
  \right).
\label{dgamma}
\end{gather}

From these relations one can obtain expressions for
$\partial(\zeta) b_{n}(\zeta)$ and $\partial(\zeta) c_{n}(\zeta)$. At
first glance the latter is nonlocal because of the last
terms in the right-hand sides of (\ref{dbeta}) and~(\ref{dgamma}).
However, this nonlocality can be quickly eliminated by means of
the operator~$\mathrm{d}/\mathrm{d}\zeta$. Indeed, using the
identities
\begin{equation}
\sum_{j,k=1}^{\infty}
\zeta^{j+k}
\sum_{l=1}^{j} u_{l} \, v_{j+k-l} =
\sum_{l=1}^{\infty} \zeta^{l} u_{l}
\sum_{j,k=1}^{\infty}
\zeta^{j+k}
v_{j+k}
\end{equation}
and
\begin{equation}
\sum_{j,k=1}^{\infty}
\zeta^{j+k}
v_{j+k} =
\sum_{m=2}^{\infty}
(m-1) \zeta^{m} v_{m} =
\left( \zeta\! {\mbox{d} \over \mbox{d}\zeta}\; - 1 \right)
\sum_{j=1}^{\infty} \zeta^{j} v_{j}
\end{equation}
one can derive from (\ref{dbeta}) and (\ref{dgamma})
\begin{gather}
  \partial(\zeta) b_{n}(\zeta) =
  b_{n}(\zeta) \left[ d_{n}(\zeta) - i \right]
  + 2 b_{n}(\zeta)\; \zeta\! {\mbox{d} \over \mbox{d}\zeta}\; d_{n}(\zeta)
  + \left[ i - 2 d_{n}(\zeta) \right] \;
    \zeta \! {\mbox{d} \over \mbox{d}\zeta} \; b_{n}(\zeta),
\label{db-zeta}
\\
  \partial(\zeta) c_{n}(\zeta) =
  c_{n}(\zeta) \left[ i - d_{n}(\zeta) \right]
  - 2 c_{n}(\zeta)\; \zeta\! {\mbox{d} \over \mbox{d}\zeta} \; d_{n}(\zeta)
  + \left[ 2 d_{n}(\zeta) - i \right] \;
    \zeta \! {\mbox{d} \over \mbox{d}\zeta} \; c_{n}(\zeta).
\label{dc-zeta}
\end{gather}
On the other hand it follows from (\ref{zcr-b-j})--(\ref{dr-j}) that
\begin{equation}
\partial(\eta) d_{n}(\xi) =
  {1 \over \xi - \eta }
  \left\{
  b_{n}(\xi) c_{n}(\eta) - c_{n}(\xi) b_{n}(\eta)
  \right\}
\label{par-d}
\end{equation}
which leads to
\begin{equation}
  \partial(\zeta) d_{n}(\zeta) =
  c_{n}(\zeta)\; {\mbox{d} \over \mbox{d}\zeta} \; b_{n}(\zeta) -
  b_{n}(\zeta)\; {\mbox{d} \over \mbox{d}\zeta} \; c_{n}(\zeta).
\label{dd-zeta}
\end{equation}
Note that expressions (\ref{db-zeta}), (\ref{dc-zeta}) and
(\ref{dd-zeta}) relate the evolutionary flows (I mean the
differentiations,
$\partial_{j}$, of the operator $\partial(\zeta)$) with the, so to
say, `auxiliary' (or `spectral') one,~$\mathrm{d}/\mathrm{d}\zeta$.
X-Mozilla-Status: 0000

Combining now (\ref{db-zeta}), (\ref{dc-zeta}) and (\ref{dd-zeta}) one can
obtain the following result:
\begin{gather}
\partial(\zeta) \; { d_{n}(\zeta) \over b_{n}(\zeta) } =
  -i \zeta\!{\mbox{d} \over \mbox{d}\zeta} \;
  { d_{n}(\zeta) \over b_{n}(\zeta) },
\\
\partial(\zeta) \; { d_{n}(\zeta) \over c_{n}(\zeta) } =
  i \zeta\!{\mbox{d} \over \mbox{d}\zeta} \;
  { d_{n}(\zeta) \over c_{n}(\zeta) }.
\end{gather}
This means that the ratio $d_{n}/b_{n}$ (resp. $d_{n}/c_{n}$) depends on
$z^{k} + i \zeta^{k}/k$ (resp. $z^{k} - i \zeta^{k}/k$). Noting also that
\begin{gather}
\lim_{\zeta \to 0} \; { d_{n}(\zeta) \over b_{n}(\zeta) } =
  {\delta^{1}_{n} \over \beta^{1}_{n} } = - q_{n},
\\
\lim_{\zeta \to 0} \; { d_{n}(\zeta) \over c_{n}(\zeta) } =
  {\delta^{1}_{n} \over \gamma^{1}_{n} } = - r_{n-1}
\end{gather}
one can present $d_{n}/b_{n}$ and $d_{n}/b_{n}$ as
\begin{gather}
  { d_{n}(z, \zeta) \over b_{n}(z, \zeta) } =
  - q_{n} ( z + i [\zeta]),
\label{db-q}
\\
  { d_{n}(z, \zeta) \over c_{n}(z, \zeta) }=
  - r_{n-1}( z - i [\zeta]),
\label{dc-r}
\end{gather}
where the designation $z + \varepsilon [\zeta]$ stands for
\begin{equation}
f(z + \varepsilon [\zeta]) =
f \left(
  z_{1} + \varepsilon \zeta,
  z_{2} + \varepsilon \zeta^{2}/2,
  z_{3} + \varepsilon \zeta^{3}/3,
  \dots
\right)
\end{equation}
and the dependence on the $\bar z_{j}$ is not indicated explicitly.
Using (\ref{db-q})--(\ref{dc-r}) one can rewrite (\ref{zcr-b})--(\ref{zcr-d}) as
\begin{gather}
  q_{n}( z + i [\zeta]) - q_{n}( z ) -
  \zeta \; { c_{n+1}(z, \zeta) \over b_{n}(z, \zeta) }
  \left[ 1 - q_{n} ( z ) r_{n} ( z - i [\zeta] ) \right]
  r_{n-1} ( z - i [\zeta]) = 0,\\
    r_{n}( z - i [\zeta]) - r_{n}( z ) -
  \zeta \; { b_{n}(z, \zeta) \over c_{n+1}(z, \zeta) }
  \left[ 1 - q_{n} ( z + i [\zeta]) r_{n} ( z ) \right]
  q_{n+1} ( z + i [\zeta]) = 0
\end{gather}
and
\begin{equation}
  b_{n}(z, \zeta)
    \left[ q_{n}( z + i [\zeta]) - q_{n} ( z ) \right] +
  c_{n+1}(z, \zeta)
    \left[ r_{n}( z ) - r_{n}( z - i [\zeta] ) \right]
  = 0,
\end{equation}
from which it follows that
\begin{gather}
q_{n}( z ) - q_{n}( z - i [\zeta] ) =
   \zeta \left[ 1 - q_{n}( z ) r_{n}( z - i [\zeta] ) \right]
   q_{n+1}( z ),
\label{fe-q}
\\
r_{n}( z ) - r_{n}( z + i [\zeta])=
   \zeta \left[ 1 - q_{n}( z + i [\zeta] ) r_{n}( z ) \right]
   r_{n-1}( z ).
\label{fe-r}
\end{gather}
These equations (which were absent in explicit form
in~\cite{1}) can be viewed as a functional representation of the `positive'
Ablowitz--Ladik subhierarchy. Expanding them in power series in $\zeta$
one can consequently obtain
\begin{gather}
  \partial_{1} q_{n} = -ip_{n}q_{n+1},
\label{dq-1}
\\
  \partial_{1} r_{n} = ip_{n}r_{n-1}
\label{dr-1}
\end{gather}
(the terms proportional to $\zeta^{1}$), then
\begin{gather}
\left( \partial_{2} - i \partial_{11} \right) q_{n} =
   2 q_{n} q_{n+1} \, \partial_{1} r_{n},
\\
\left( \partial_{2} + i \partial_{11} \right) r_{n} =
   2 r_{n-1} r_{n} \, \partial_{1} q_{n}
\end{gather}
(the terms proportional to $\zeta^{2}$), which using equations
(\ref{dq-1}), (\ref{dr-1}) can be transformed to
\begin{gather}
\partial_{2} q_{n}=
  ir_{n-1}p_{n}q_{n}q_{n+1}+ip_{n}r_{n}q_{n+1}^{2}-ip_{n}p_{n+1}q_{n+2},
\label{dq-2}
\\
\partial_{2} r_{n} =
  ir_{n-2}p_{n-1}p_{n}-ir_{n-1}^{2}p_{n}q_{n}-ir_{n-1}p_{n}r_{n}q_{n+1}
\label{dr-2}
\end{gather}
and all other equations in the similar way.

The functional equations for the `negative' subhierarchy can be
written as
\begin{gather}
q_{n}( \bar z )  - q_{n}\left( \bar z - i \left[\zeta^{-1}\right] \right) =
   \zeta^{-1}
   \left[ 1 - q_{n}( \bar z ) r_{n}\left( \bar z - i \left[\zeta^{-1}\right] \right) \right]
   q_{n-1}( \bar z ),
\label{bfe-q}
\\
r_{n}( \bar z ) - r_{n}\left( \bar z + i \left[\zeta^{-1}\right]\right) =
   \zeta^{-1}
   \left[ 1 - q_{n}\left( \bar z + i \left[\zeta^{-1}\right] \right) r_{n}( \bar z ) \right]
   r_{n+1}( \bar z )
\label{bfe-r}
\end{gather}
with the dependence on the $z_{k}$ being omitted.

\section{Functional equations for tau-functions} \label{sec-tau}

The functional equations obtained in the previous section, strictly
speaking, complete the solution of the problem formulated in the title  of
this paper: equations (\ref{fe-q}), (\ref{fe-r}) and~(\ref{bfe-q}),
(\ref{bfe-r}) are sufficient to generate all equations of the ALH.
However, we proceed further and present some results which
seem to be interesting from various viewpoints.

It is widely known that intrinsic properties of integrable equations and
hierarchies become more transparent when one uses Hirota's bilinear
representation, i.e.\ when one operates not with the solutions themselves
(the $q_{n}$s and $r_{n}$s in our case) but with the so-called
tau-functions. The tau-functions of the ALH are defined by
\begin{equation}
  p_{n} = { \tau_{n-1} \tau_{n+1} \over \tau_{n}^{2} },
\qquad
  q_{n} = { \sigma_{n} \over \tau_{n} },
\qquad
  r_{n} = { \rho_{n} \over \tau_{n} }.
\label{tau-def}
\end{equation}
The definition of $p_{n}$, $p_{n}=1-q_{n}r_{n}$, rewritten in terms of tau
functions, leads to the following relation between them:
\begin{equation}
  \tau_{n-1} \tau_{n+1} = \tau_{n}^{2} - \sigma_{n} \rho_{n}.
\label{tau-pqr}
\end{equation}
Our aim now is to derive the functional equations for
$\tau_{n}$, $\sigma_{n}$ and $\rho_{n}$ using equations obtained in the previous
section. To this end we rewrite (\ref{fe-q}) and (\ref{fe-r}) as
\begin{gather}
  q_{n}^{+} - q_{n} = \zeta X_{n} q_{n+1}^{+},
\label{fe1-q}
\\
  r_{n} - r_{n}^{+}= \zeta X_{n} r_{n-1}
\label{fe1-r}
\end{gather}
with
\begin{equation}
X_{n} = 1 - q_{n}^{+} r_{n},
\label{tau-x-def}
\end{equation}
where the designation
\begin{equation}
f_{n}^{+} \equiv f_{n}( z + i [\zeta], \bar z )
\end{equation}
is used. It follows from (\ref{fe1-q}) and (\ref{fe1-r}) that the quantity
$X_{n}$ satisfies the identities
\begin{gather}
X_{n}=
   1 - r_{n} \left( q_{n} + \zeta X_{n} q_{n+1}^{+} \right)
   =
   p_{n} - \zeta r_{n} q_{n+1}^{+} X_{n}
\\ \phantom{X_{n}} {}=
   1 - q_{n}^{+} \left( r_{n}^{+} + \zeta X_{n} r_{n-1} \right)
   =
   p_{n}^{+} - \zeta r_{n-1} q_{n}^{+} X_{n}
\end{gather}
which yields
\begin{equation}
X_{n}
=
{ p_{n} \over 1 + \zeta r_{n} q_{n+1}^{+} }
=
{ p_{n}^{+} \over 1 + \zeta r_{n-1} q_{n}^{+} }
\end{equation}
and
\begin{equation}
{ X_{n} \over X_{n+1} } =
{ p_{n} \over p_{n+1}^{+} } =
{ \tau_{n-1} \tau_{n+1}^{+} \over \tau_{n} \tau_{n}^{+} }
\cdot
{ \tau_{n+1} \tau_{n+1}^{+} \over \tau_{n} \tau_{n+2}^{+} }.
\end{equation}

From the last equation one can easily obtain that
\begin{equation}
X_{n} = x { \tau_{n-1} \tau_{n+1}^{+} \over \tau_{n} \tau_{n}^{+} },
\qquad
  x = \mbox{constant}.
\label{tau-x-sol}
\end{equation}
In what follows we restrict our attention to the case
\begin{equation}
x = 1
\end{equation}
which corresponds to the so-called `finite-density' boundary conditions
\begin{equation}
\lim_{n \to \pm\infty} p_{n} = \mbox{constant}
\end{equation}
(which also includes the vanishing boundary conditions considered
in~\cite{1}). In some other important cases, first of all the case
when $x=x(\zeta)$ depends on $\zeta$ but does not depend on the
$z_{j}$s (such constant $x$ appears in the quasi-periodic
situation), one can construct the substitution $(\tau_{n},
\sigma_{n}, \rho_{n}) \to \exp\{ \alpha n^{2} + n f_{1}( z ) +
f_{0}( z ) \} (\tau_{n}, \sigma_{n}, \rho_{n}) $, where $f_{1,0}(
z )$ are some linear function of the $z_{j}$s, which eliminates
the constant $x(\zeta)$ and hence provides reduction to the case
of this paper.
X-Mozilla-Status: 0000

Substituting (\ref{tau-x-sol}) in (\ref{tau-x-def}) and then in
(\ref{fe1-q}), (\ref{fe1-r}) one can obtain the equations
\begin{gather}
\tau_{n}  ( z ) \; \tau_{n}  ( z + i[\zeta] ) -
\rho_{n}  ( z ) \; \sigma_{n}( z + i[\zeta] ) =
\phantom{\zeta} \;
  \tau_{n-1}( z ) \; \tau_{n+1}( z + i[\zeta] ),
\label{tau-fe-t}
\\
\tau_{n}  ( z ) \; \sigma_{n}( z + i[\zeta] ) -
\sigma_{n}( z ) \; \tau_{n}  ( z + i[\zeta] ) =
\zeta \;
  \tau_{n-1}( z ) \; \sigma_{n+1}( z + i[\zeta] ),
\label{tau-fe-s}
\\
\rho_{n}  ( z ) \; \tau_{n}  ( z + i[\zeta] ) -
\tau_{n}  ( z ) \; \rho_{n}  ( z + i[\zeta] ) =
\zeta  \;
\rho_{n-1}( z ) \; \tau_{n+1}( z + i[\zeta] ),
\label{tau-fe-r}
\end{gather}
where dependence on the $\bar z_{j}$s is omitted. This system was
the main result of the paper~\cite{1}. In what follows we use some
other form of (\ref{tau-fe-t})--(\ref{tau-fe-r}), which can be
obtained from the latter by means of~(\ref{tau-pqr}):
\begin{gather}
\tau_{n+1}( z ) \; \tau_{n}( z + i[\zeta] ) -
\tau_{n}  ( z ) \; \tau_{n+1}( z + i[\zeta] ) =
  \zeta \; \rho_{n}( z ) \; \sigma_{n+1}( z + i[\zeta] ),
\label{tau-fe1-t}
\\
\tau_{n+1}( z ) \; \sigma_{n}( z + i[\zeta] ) -
\sigma_{n}( z ) \; \tau_{n+1}( z + i[\zeta] ) =
  \zeta \; \tau_{n}( z ) \; \sigma_{n+1}( z + i[\zeta] ),
\label{tau-fe1-s}
\\
\rho_{n+1}( z ) \; \tau_{n}  ( z + i[\zeta] ) -
\tau_{n}  ( z ) \; \rho_{n+1}( z + i[\zeta] ) =
  \zeta \; \rho_{n}( z ) \; \tau_{n+1}( z + i[\zeta] ).
\label{tau-fe1-r}
\end{gather}

Before proceeding further we present, without derivation, the
equations which are equivalent to the `negative' subhierarchy. The systems
analogous to (\ref{tau-fe-t})--(\ref{tau-fe-r}) and (\ref{tau-fe1-t})--(\ref{tau-fe1-r})
in the `negative' case can be written as
\begin{gather}
\tau_{n} ( \bar z ) \; \tau_{n} \left( \bar z + i\left[\zeta^{-1}\right]\right) -
\rho_{n} ( \bar z ) \; \sigma_{n}\left( \bar z + i\left[\zeta^{-1}\right]\right)\nonumber\\
\qquad{} =
  \tau_{n+1}( \bar z ) \; \tau_{n-1}\left( \bar z + i\left[\zeta^{-1}\right]\right ),
\label{tau-nfe-t}
\\
\tau_{n}  ( \bar z ) \; \sigma_{n}\left( \bar z + i\left[\zeta^{-1}\right]\right) -
\sigma_{n}( \bar z ) \; \tau_{n}  \left( \bar z + i\left[\zeta^{-1}\right]\right) \nonumber\\
\qquad{}=
\zeta^{-1} \;
  \tau_{n+1}( \bar z ) \; \sigma_{n-1}\left( \bar z + i\left[\zeta^{-1}\right]\right),
\label{tau-nfe-s}
\\
\rho_{n}  ( \bar z ) \; \tau_{n}  \left( \bar z + i\left[\zeta^{-1}\right]\right) -
\tau_{n}  ( \bar z ) \; \rho_{n}  \left( \bar z + i\left[\zeta^{-1}\right]\right)\nonumber\\
\qquad{}=
\zeta^{-1}  \;
  \rho_{n+1}( \bar z ) \; \tau_{n-1}\left( \bar z + i\left[\zeta^{-1}\right]\right)
\label{tau-nfe-r}
\end{gather}
and
\begin{gather}
  \tau_{n-1}( \bar z ) \; \tau_{n}\left( \bar z + i\left[\zeta^{-1}\right]\right) -
  \tau_{n} ( \bar z ) \; \tau_{n-1}\left( \bar z + i\left[\zeta^{-1}\right]\right)\nonumber\\
\qquad{}
=
    \zeta^{-1} \;
    \rho_{n}( \bar z ) \; \sigma_{n-1}\left( \bar z + i\left[\zeta^{-1}\right]\right ),
  \label{tau-nfe1-t}
\\
  \tau_{n-1}( \bar z ) \; \sigma_{n}\left( \bar z + i\left[\zeta^{-1}\right]\right) -
  \sigma_{n}( \bar z ) \; \tau_{n-1}\left( \bar z + i\left[\zeta^{-1}\right]\right)\nonumber\\
\qquad{}=
    \zeta^{-1} \;
    \tau_{n}( \bar z ) \; \sigma_{n-1}\left( \bar z + i\left[\zeta^{-1}\right]\right),
\label{tau-nfe1-s}
\\
  \rho_{n-1}( \bar z ) \; \tau_{n}  \left( \bar z + i\left[\zeta^{-1}\right]\right) -
  \tau_{n}  ( \bar z ) \; \rho_{n-1}\left( \bar z + i\left[\zeta^{-1}\right]\right)\nonumber\\
\qquad{}  =
    \zeta^{-1} \;
    \rho_{n}( \bar z ) \; \tau_{n-1}\left( \bar z + i\left[\zeta^{-1}\right]\right).
\label{tau-nfe1-r}
\end{gather}

Returning to the `positive' subhierarchy, we note that equations
(\ref{tau-fe-t})--(\ref{tau-fe-r}) contain much information
about the ALH and we now derive some of their consequences.
Firstly we continue discussion of the zero-curvature
representation of the ALH started in Section~\ref{sec-zcr}. The
relation~(\ref{zcr-stat}) indicates that the matrix
$\mathcal{V}_{n}$ can be presented in the form
$\mathcal{V}_{n}=\Phi_{n} C \Phi_{n}^{-1}$, where $C$ is a
constant (with respect to~$n$) matrix and $\Phi_{n}$ is a solution
of the scattering problem~(\ref{zcr-sp}). Thus one can hope to
obtain from (\ref{calv-def}) some, at least formal, solution of
the scattering problem, which can be then modified to become a
solution of the evolutionary problem~(\ref{zcr-evol-pos}) as well,
i.e.\ to obtain the Baker--Akhiezer function of our problem. One can
hardly derive a closed local expression for $\Phi_{n}$ in terms of
the quantities $b_{n}$, $c_{n}$ and $d_{n}$, but this can be done
in terms of the tau-functions.

Consider the matrix
\begin{equation}
F_{n}(z, \bar z, \lambda) =
{ 1 \over \tau_{n-1} (z, \bar z) }
\begin{pmatrix}
           \tau_{n}   \left(z + i \left[\lambda^{2}\right], \bar z\right)  &
   \lambda \rho_{n-1} \left(z - i \left[\lambda^{2}\right], \bar z\right) \vspace{1mm}\\
  -\lambda \sigma_{n} \left(z + i \left[\lambda^{2}\right], \bar z\right) &
           \tau_{n-1} \left(z - i \left[\lambda^{2}\right], \bar z\right)
\end{pmatrix}.
\end{equation}
Using (\ref{tau-fe-t})--(\ref{tau-fe1-r}) one can straightforwardly
verify that this matrix satisfies the equation
\begin{equation}
U_{n} (\lambda ) \; F_{n}(\lambda) =
F_{n+1}(\lambda)   \begin{pmatrix} \lambda & 0 \\ 0 & \lambda^{-1}
\end{pmatrix}
\label{uf}
\end{equation}
which means that
\begin{equation}
\Phi_{n} (z, \bar z, \lambda ) =
F_{n}(z, \bar z, \lambda)
\begin{pmatrix} \lambda^{n} & 0 \\ 0 & \lambda^{-n}
\end{pmatrix}
\label{phi-f}
\end{equation}
is a (formal) solution of (\ref{zcr-sp}).

Analogously from the `negative' subhierarchy one can derive another
solution of (\ref{uf}), $\bar F_{n}$,
\begin{equation}
\bar F_{n}(z, \bar z,\lambda) =
{ 1 \over \tau_{n-1} (z, \bar z) }
\begin{pmatrix}
                 \tau_{n-1}   \left(z, \bar z + i \left[\lambda^{-2}\right]\right)\! &
   -\lambda^{-1} \rho_{n}     \left(z, \bar z - i \left[\lambda^{-2}\right]\right) \\
    \lambda^{-1} \sigma_{n-1} \left(z, \bar z + i \left[\lambda^{-2}\right]\right)\! &
                 \tau_{n}     \left(z, \bar z - i \left[\lambda^{-2}\right]\right)
\end{pmatrix}\!
\end{equation}
and, hence, another solution of the scattering problem:
\begin{equation}
  \bar\Phi_{n} (z, \bar z, \lambda ) =
  \bar F_{n}(z, \bar z, \lambda)
  \begin{pmatrix} \lambda^{n} & 0 \\ 0 & \lambda^{-n}
\end{pmatrix}.
\end{equation}

\section{Discrete-time Ablowitz--Ladik equations} \label{sec-suris}

In Section \ref{sec-fe} we have derived the functional equations
(\ref{fe-q}), (\ref{fe-r}) and (\ref{bfe-q}), (\ref{bfe-r}) which
we now rewrite as
\begin{gather}
\widehat q_{n} - q_{n} =
   \xi \left[ 1 - \widehat q_{n} r_{n} \right]
   \widehat q_{n+1},
\label{su-q}
\\
r_{n} - \widehat r_{n} =
   \xi \left[ 1 - \widehat q_{n} r_{n} \right]
   r_{n-1},
\label{su-r}
\end{gather}
where $\widehat q_{n} = q_{n} (z + i [\xi], \bar z)$ etc and
\begin{gather}
q_{n} - \widetilde q_{n} =
   \xi^{-1} \left[ 1 - q_{n} \widetilde r_{n} \right]
   q_{n-1},
\label{bsu-q}
\\
\widetilde r_{n} - r_{n} =
   \xi^{-1} \left[ 1 - q_{n} \widetilde r_{n} \right]
   \widetilde r_{n+1}
\label{bsu-r}
\end{gather}
with $\widetilde q_{n} = q_{n} (z, \bar z - i [\xi^{-1}])$ etc,
and demonstrated that these equations are equivalent to the ALH.
It should be noted that equations (\ref{su-q})--(\ref{bsu-r})
have already been discussed in the literature in some other
context. We mean the so-called discrete-time Ablowitz--Ladik
equations, which arise as compatibility conditions for the
discrete-discrete linear system
\begin{equation}
  \Psi_{n+1} = U_{n} \Psi_{n} \label{dd-sp}
\end{equation}
and
\begin{equation}
  \widehat \Psi_{n} = M_{n} \Psi_{n}, \label{dd-evol}
\end{equation}
where (\ref{dd-sp}) is the scattering problem of the ALH, and
which are the integrable discretizations of the
differential-difference Ablowitz--Ladik equations.

Before proceeding further, we note that equations
(\ref{dd-sp}) and (\ref{dd-evol}) are essentially discrete equations.
Strictly speaking they should be written as
\begin{equation}
\Psi_{n+1,m} = U_{n,m} \Psi_{n,m}
\qquad
\mbox{and}
\qquad
\Psi_{n,m+1} = M_{n,m} \Psi_{n,m}
\end{equation}
and the resulting equations will take the form
\begin{gather}
  q_{n,m} - q_{n,m+1} =
    Q\left(q_{n,m},r_{n,m},q_{n \pm 1, m \pm 1},r_{n \pm 1,m \pm 1}, \ldots \right),
\label{dd-q}
\\
  r_{n,m} - r_{n,m+1} =
    R\left(q_{n,m},r_{n,m},q_{n \pm 1, m \pm 1},r_{n \pm 1,m \pm 1}, \ldots \right).
\label{dd-r}
\end{gather}
Depending upon the choice of the $M$-matrix one can derive
equations which cannot be reduced by some limiting procedure to
the evolutionary equations of the ALH (for example, the system
(\ref{dd-sp}) and (\ref{dd-evol}) is the starting point for
obtaining B\"acklund transformations). However, among equations
(\ref{dd-q}) and (\ref{dd-r}) there is a big class of equations
having the continuous limit, and we discuss here these equations,
which were studied in~\cite{AL3,Suris}. In the paper~\cite{Suris}
by Suris it has been shown that the nonlocal scheme of Ablowitz
and Ladik~\cite{AL3} can be factorized into the product of the
local ones which are exactly (up to redefinition of variables)
equations~(\ref{su-q}), (\ref{su-r}) and~(\ref{bsu-q}),
(\ref{bsu-r}). Thus the results of the previous sections lead to
a, so to say, explicit relation between the ALH and the equations
of~\cite{Suris} in terms of the solutions: each solution of, for
example, the `positive' subhierarchy provides a solution of the
discrete-time equations (\ref{su-q}) and (\ref{su-r}) after
identifying $\widehat q_{n}$ and $\widehat r_{n}$ with $q_{n} (z +
i [\xi], \bar z)$ and $r_{n} (z + i [\xi], \bar z)$. In other
words the discrete flow is equivalent to (or at least can be
modeled as) a properly chosen combination of the infinite number
of continuous ones: the shift $\Psi_{n} \to \widehat\Psi_{n}$ can
be obtained as simultaneous shifts in all $z_{k}$-directions,
$z_{k} \to z_{k} + i\xi^{k}/k$,  $(k=1,2,\ldots)$.

This result is not new. It has been already obtained by Suris
in~\cite{Suris} in the framework of the Hamiltonian approach. One can
find in \cite{Suris} an explicit expression for the Hamiltonian of
the discrete-time flow, from which it follows that it can be
presented as a linear combination of the Hamiltonians of the
differential Ablowitz--Ladik equations (Proposition~12
of~\cite{Suris}). From this standpoint the above consideration can be
viewed as an alternative derivation of this result, which does not
exploit Hamiltonian structures, uses only local properties of the
ALH and, hence, is not so sensitive to the boundary conditions.

The main subject of this paper is the ALH, not
equations~(\ref{dd-q}) and (\ref{dd-r}) themselves. For us the representation
(\ref{su-q})--(\ref{bsu-r}) is a way to obtain some results for
the Ablowitz--Ladik equations by studying the discrete system which
is, in some sense, a more simple object than the original system of
differential equations. In~\cite{qps} it was shown how this
approach can be used to derive, without much effort, the
finite-genus quasiperiodical solutions, the problem which is
rather difficult technically if one solves it in the language of
differential equations (see~\cite{MEKL}). Firstly one solves the
discrete equations (\ref{su-q})--(\ref{bsu-r}) using Fay's
identities for the $\theta$-functions and then adjusts some
constants in such a way that the discrete shifts become Miwa's
shifts $z_{k} \to z_{k} + i\xi^{k}/k$,
$\bar z_{k} \to \bar z_{k} - i\xi^{-k}/k$.

In what follows we use the discrete time representation
(\ref{dd-sp}) and (\ref{dd-evol}) to obtain the `superposition'
formulae for the tau-functions of the ALH. Thus we want to
finish this section by writing down the $M$-matrices which lead to
(\ref{su-q})--(\ref{bsu-r}).

The matrix $M_{n}$ describing shifts in the `positive' discrete
direction, ($z \to z + i[\xi]$) can be written as
\begin{equation}
M_{n}(z,\lambda; \xi) = { 1 \over \tau_{n} \widehat\tau_{n-1} }
\begin{pmatrix}
  \tau_{n} \widehat\tau_{n-1} -
    \lambda^{-2} \xi \tau_{n-1} \widehat\tau_{n}    &
  \lambda^{-1} \xi \, \rho_{n-1} \widehat\tau_{n}   \vspace{1mm}\\
  \lambda^{-1} \xi \, \tau_{n-1} \widehat\sigma_{n} &
  \tau_{n-1} \widehat\tau_{n}
\end{pmatrix}.
\label{m-matrix}
\end{equation}
Using the functional equations (\ref{tau-fe-t})--(\ref{tau-fe1-r})
 for the quantities $\widehat\tau_{n} =
\tau_{n}(z+i[\xi])$ etc, one can verify the relation
\begin{equation}
  M_{n+1}( \lambda; \xi) \; U_{n} ( \lambda ) =
  \widehat U_{n} ( \lambda ) \; M_{n}( \lambda; \xi ),
\label{dd-zcr}
\end{equation}
where
\begin{equation}
  \widehat U_{n}( \lambda ) =
  \begin{pmatrix}
    \lambda & \widehat r_{n} \\
    \widehat q_{n} & \lambda^{-1}
\end{pmatrix}.
\end{equation}
Analogously the `negative' discrete-time shift matrices, $\bar
M_{n}$, can be presented as
\begin{equation}
\bar M_{n}(\bar z,\lambda; \xi) = { 1 \over \tau_{n}
\widetilde\tau_{n-1} }
 \begin{pmatrix}
  \tau_{n-1} \widetilde\tau_{n}                         &
  \lambda \xi^{-1} \, \tau_{n-1} \widetilde\rho_{n}   \vspace{1mm}\\
  \lambda \xi^{-1} \, \sigma_{n-1} \widetilde\tau_{n} &
  \tau_{n} \widetilde\tau_{n-1} -
     \lambda^{2} \xi^{-1} \, \tau_{n-1} \widetilde\tau_{n}
\end{pmatrix} \label{bm-matrix}
\end{equation}
with $\widetilde\tau_{n}$ standing for $\tau_{n}\left(\bar z -
i\left[\xi^{-1}\right]\right)$ etc. These matrices after substitution into
$\bar M_{n+1} U_{n} = \widetilde U_{n} \bar M_{n}$ lead to
equations (\ref{bsu-q})--(\ref{bsu-r}).

\section{Fay's identities for the tau-functions} \label{sec-fay}

The aim of this section is to derive some results for the ALH by
means of, so to say, discrete-time approach. We do not use the
evolutionary part of the ZCR, (\ref{zcr-evol}), and explicit form
of the ALH flows. All we need now is the fact that their combined
action can be presented as the map discussed in the previous
section, and can be described by the
$M$-matrices (\ref{m-matrix}) or (\ref{bm-matrix}). The question,
which we address now, is the superposition of the maps $q_{n} \to
\hat q_{n}$ (or, in other words, superposition of Miwa's shifts) which
enables us to obtain the so-called Fay's identities for the ALH.

The discrete-time zero-curvature condition (\ref{dd-zcr}), when
considered in the framework of the functional equations for the
ALH, is
\begin{equation}
M_{n+1}( z, \lambda; \xi) \;     U_{n} ( z, \lambda ) =
U_{n} ( z + i[\xi], \lambda ) \; M_{n}( z, \lambda; \xi)
\end{equation}
(the dependence on $\bar z=(\bar z_{1},\bar z_{2},\ldots)$ is omitted).
Consider now the matrices (\ref{phi-f}) which solve
\begin{equation}
  \Phi_{n+1} = U_{n} \Phi_{n}.
\end{equation}
Acting on this equation by $M_{n+1}$ one can obtain
\begin{equation}
  M_{n+1}\Phi_{n+1} = \widehat U_{n} \, M_{n}\Phi_{n},
\qquad
  \widehat U_{n} \equiv U_{n} ( z + i[\xi], \lambda ).
\end{equation}
Thus the product $M_{n}\Phi_{n}$ differs from $\widehat\Phi_{n}$ only by
constant matrix $C$ or, in terms of $F_{n}$'s,
\begin{equation}
  M_{n} F_{n} = \widehat F_{n} \Lambda^{n} C \Lambda^{-n},
\qquad
  \Lambda =  \begin{pmatrix}\lambda & 0 \\ 0 & \lambda^{-1}
\end{pmatrix}.
\end{equation}
By direct calculation one can verify that the nondiagonal elements of
the matrix $\widehat F_{n}^{-1} M_{n} F_{n}$ are zero,
\begin{gather}
  \left[
  F_{n}^{-1} ( z + i[\xi], \lambda ) \;
  M_{n}( z, \lambda; \xi)            \;
  F_{n} ( z, \lambda )
  \right]^{(12)}
  = 0,
\\
  \left[
  F_{n}^{-1} ( z + i[\xi], \lambda ) \;
  M_{n}( z, \lambda; \xi) \; F_{n} ( z, \lambda ) \right]^{(21)}
  = 0,
\end{gather}
i.e.\ $C$ is a diagonal matrix, which hence commutes with $\Lambda$,
\begin{equation}
M_{n}( z, \lambda; \xi) \; F_{n} ( z, \lambda ) =
F_{n} ( z + i[\xi], \lambda ) \;
\begin{pmatrix} C_{1}(\zeta; \xi) & 0 \\ 0 & C_{2}(\zeta; \xi)
\end{pmatrix},
\qquad
  \zeta = \lambda^{2}.
\end{equation}
The upper-left element of this matrix relation can be written as
\begin{gather}
 \kappa(\zeta, \xi) \,
  \tau_{n-1}( z )  \, \tau_{n}  ( z + i[\zeta] + i [\xi] )\nonumber\\
\label{two-shift-t}
\qquad{} =
  \zeta \,
    \tau_{n} ( z + i[\zeta] ) \,  \tau_{n-1} ( z + i[\xi] ) -
  \xi \,
    \tau_{n} ( z + i[\xi] ) \, \tau_{n-1} ( z + i[\zeta] ),
\end{gather}
where
\begin{equation}
\kappa\left( \zeta, \xi \right) =
  -\kappa\left( \xi, \zeta \right) =
  \zeta C_{1}(\zeta; \xi).
\end{equation}
It can be shown that analogous identities can be derived for all
tau-functions,
\begin{gather}
\kappa(\zeta, \xi) \,
  \omega_{n-1}( z )  \, \omega_{n}  ( z + i[\zeta] + i [\xi] )\nonumber\\
\label{two-shift-w}
\qquad {}=
\zeta \, \omega_{n} ( z + i[\zeta] ) \, \omega_{n-1} ( z + i[\xi] ) -
\xi   \, \omega_{n} ( z + i[\xi] )   \, \omega_{n-1} ( z + i[\zeta] ),
\end{gather}
where
\begin{equation}
\omega_{n} = \sigma_{n}, \quad \rho_{n} \quad \mbox{or} \quad \tau_{n}.
\end{equation}

These equations are the simplest `determinant' identities for the
`positive' ALH. They can be generalized to obtain determinantal
representation of superposition of arbitrary number of Miwa's shifts:
\begin{gather}
K \left( \zeta_{1}, \ldots , \zeta_{M} \right) \,
\omega_{n}(z)
\cdots
\omega_{n+M-2}( z  ) \,
\omega_{n+M-1}( z + i[\zeta_{1}] + \cdots + i[\zeta_{M}] )\nonumber\\
\label{det-w}
\qquad {}=
\det \left|
\zeta_{r}^{s-1} \omega_{n+s-1}( z + i[\zeta_{r}] )
\right|_{r,s=1, \ldots ,M},
\end{gather}
where
\begin{equation}
K \left( \zeta_{1}, \ldots, \zeta_{M} \right) =
\prod_{1 \leq j < k \leq M}
   \kappa\left( \zeta_{k}, \,  \zeta_{j} \right).
\end{equation}
Expanding the determinants (say, over the first few columns) one
can obtain a large number of multilinear relations the simplest of
which (one-column) are
\begin{gather}
\omega_{n-M+1}(z)    \,
\omega_{n}( z + i[\zeta_{1}] + \cdots  + i[\zeta_{M}] ) \nonumber\\
\qquad{}
=
\sum_{j=1}^{M}
    K_{j}\,
  \omega_{n-M+1}\left( z + i[\zeta_{j}] )
  \,
  \omega_{n}( z + i[\zeta_{1}] + \cdots
            + \widehat{ i[\zeta_{j}] } + \cdots
            + i[\zeta_{M}]\right).
\end{gather}
The symbol $\; \widehat{ } \;$ here indicates that the
corresponding quantity should be omitted and
\begin{equation}
  K_{j} =
  {\prod_{k \ne j}}^{'}
  { \zeta_{k}  \over \kappa(\zeta_{k}, \zeta_{j}) }.
\end{equation}
Clearly analogous formulae can be derived for the `negative'
shifts as well.

\section{Integrable discretizations of higher flows\\ and `reduced' Miwa's shifts}
 \label{sec-hdf}

After we have derived the superposition formulae for Miwa's
shifts, we can address the following question (which has been
raised by the referee of this paper). The discrete-time equation
discussed in the Section~\ref{sec-suris} has been considered by
Suris as the integrable discretization of the {\em simplest}
Ablowitz-Ladik flows. At the same time it interpolates {\em all}
continuous ALH flows. However it is possible to construct the
integrable discretizations of the {\em higher} Ablowitz--Ladik
flows as has been done, e.g., in the paper~\cite{Suris2}. So the
question is how these higher discrete-time equations can be
related to the original ALH and to its Miwa's shift
representation?

To get some insight into this problem consider the twice shifted
tau-functions,
\begin{equation}
  \tilde\omega_{n} = \omega_{n} ( z + i[\xi] + i[\eta] ),
  \qquad
  \omega_{n} = \sigma_{n},\quad \rho_{n} \quad \mbox{or} \quad \tau_{n}.
\end{equation}
One can derive from formulae (\ref{two-shift-w}) the following
identity
\begin{equation}
  \xi\eta \left(
    \rho_{n} \tilde\varphi_{n+1} - \tau_{n-1} \tilde\sigma_{n+2}
  \right) +
  \left( \xi + \eta \right) \tau_{n} \tilde\sigma_{n+1} +
  \sigma_{n} \tilde\tau_{n+1} - \tau_{n+1} \tilde\sigma_{n}
  = 0,
\label{two-shifts}
\end{equation}
where $\varphi_{n}$ is defined by
\begin{equation}
  \tau_{n} \varphi_{n} =
  \sigma_{n}^{2} - \sigma_{n-1} \sigma_{n+1}
\end{equation}
(in other words $\varphi_{n}$ is the next tau-function in the
chain $\cdots \to \rho_{n} \to \tau_{n} \to \sigma_{n} \to
\varphi_{n} \to \cdots$, compare with (\ref{tau-pqr})). After the
imposition of the restriction
\begin{equation}
  \eta = - \xi
\end{equation}
the relation (\ref{two-shifts}) leads to
\begin{equation}
 { \tilde q_{n} - q_{n} \over \xi^{2} } =
  P_{n} \left(
    \tilde p_{n+1} \tilde q_{n+2}
    - r_{n} \tilde q_{n+1}^{2}
    - r_{n-1} q_{n} \tilde q_{n+1}
  - \xi^{2} r_{n-1} P_{n} \tilde p_{n+1} \tilde q_{n+1} \tilde q_{n+2}
  \right)
\label{dte-two-q}
\end{equation}
and
\begin{equation}
  { r_{n} - \tilde r_{n} \over \xi^{2} } =
  P_{n} \left(
    r_{n-2} p_{n-1}
    - r_{n-1}^{2} \tilde q_{n}
    - r_{n-1} \tilde r_{n} \tilde q_{n+1}
  - \xi^{2} r_{n-2} r_{n-1} p_{n-1} P_{n} \tilde q_{n+1}
  \right),
\label{dte-two-r}
\end{equation}
where
\begin{equation}
  P_{n} = P_{n}(\xi) =
  { 1 - r_{n} \tilde q_{n}
    \over
    1 + \xi^{2} r_{n-1} \tilde q_{n+1}
  }.
\end{equation}
If one considers $q \to \tilde q$ as the discrete-time shift, then
equations (\ref{dte-two-q}) and (\ref{dte-two-r}) form a~closed
discrete-time system.  On the other hand
\begin{equation}
  \tilde q, \tilde r =
  q,r \left(
    z_{1}, z_{2} + i\xi^{2}, \ldots
  \right)
\end{equation}
and in the $\xi \to 0$ limit the left-hand sides of
(\ref{dte-two-q}) and (\ref{dte-two-r}) become $i \partial q /
\partial z_{2}$ and $i \partial r / \partial z_{2}$, while the
right-hand sides become the expressions for the second positive
ALH flow, see~(\ref{dq-2}) and~(\ref{dr-2}). Thus the system~(\ref{dte-two-q})
and~(\ref{dte-two-r}) can be viewed as an
integrable discretization of the second ALH flow and, indeed, this
system is a condensed form of the corresponding equations of the
work~\cite{Suris2}.

In other words the second discrete-time flow can be constructed
using the product of two Miwa's shifts $T_{z}(\xi) T_{z}(-\xi)$,
where
\begin{equation}
  T_{z}(\xi): \qquad z_{k} \to z_{k} + i \xi^{k} / k.
\end{equation}
Looking once more on the structure of the double shifted functions
\begin{equation}
  \tilde q,\tilde r =
  q,r \left(
    z_{1}, z_{2} + i\xi^{2},
    z_{3}, z_{4} + i\xi^{2}/2, \ldots
  \right)
\end{equation}
we easily note that the odd $z_{k}$'s remain unchanged, while
the even variables gain shifts proportional to powers of
$\xi^{2}$,
\begin{equation}
  T_{z}^{(2)}(\xi) =
  T_{z}(\xi)T_{z}(-\xi): \qquad
  \left\{
  \begin{array}{l}
  z_{2k-1}  \to z_{2k-1} ,
  \vspace{1mm}\\
  z_{2k}    \to  z_{2k} + i \xi^{2k} / k.
  \end{array}
  \right.
\end{equation}
So, if one forgets for a while about the odd $z_{k}$ and
introduces ``even" variables, $y_{k}$, by $y_{k}=z_{2k}$, then
the action of the binary Miwa's shift $T_{z}^{(2)}(\xi)$ can be
presented as the action of the usual Miwa's shift with squared
parameter,
\begin{equation}
  T_{y}(\xi^{2}): \qquad
  y_{k} \to y_{k} + i \left(\xi^{2}\right)^{k} / k.
\end{equation}
Thus we can consider $T_{z}(\xi)T_{z}(-\xi)$ as some kind of
``reduced'' Miwa's shift, which generates the second discrete-time
flow.

This construction can be extended. Introducing the $M$th-order
product of $T_{z}$,
\begin{equation}
  T_{z}^{(M)}(\xi) =
  \prod_{\ell=0}^{M-1}
    T_{z}\left( \epsilon^{\ell}\xi \right),
  \qquad
  \epsilon^{M} = 1,
\end{equation}
one can easily conclude that this operator affects only each $M$th
variable,
\begin{equation}
  T_{z}^{(M)}(\xi): \qquad
  \left\{
  \begin{array}{l}
  z_{k}  \to  z_{k} \qquad \mathrm{for} \quad k \ne Ml,\vspace{1mm}\\
  z_{Ml}   \to z_{Ml} + i \xi^{Ml} / l.
  \end{array}
  \right.
\end{equation}
So, if we construct equations similar to (\ref{dte-two-q}) and
(\ref{dte-two-r}),
\begin{equation}
  { \tilde q_{n} - q_{n} \over \xi^{M} } =  \cdots
  \qquad
  \mathrm{and}
  \qquad
  { r_{n} - \tilde r_{n} \over \xi^{M} } =  \cdots,
  \label{dte-M-qr}
\end{equation}
where $ \tilde q_{n}(z) = T_{z}^{(M)}(\xi) \, q_{n}(z)$, $ \tilde
r_{n}(z) = T_{z}^{(M)}(\xi) \, r_{n}(z)$, then in the
$\xi \to 0$ limit they become the $M$th Ablowitz--Ladik equation.
Thus the $M$th discrete-time flow is produced by
$T_{z}^{(M)}(\xi)$ which after introducing variables
\begin{equation}
  y_{m}^{(M)} = z_{Mm}
\end{equation}
can be rewritten as the usual Miwa's shift with the $M$th power of
the parameter:
\begin{equation}
  T_{y^{(M)}}\left(\xi^{M}\right): \qquad
  y_{k} \to y_{k} + i \left(\xi^{M}\right)^{k} / k.
\end{equation}

However, to construct equations of the type (\ref{dte-M-qr})
starting from the superposition formulae (\ref{two-shift-w}) is a
rather cumbersome procedure and, if one wants to derive the
$M$th discrete-time Ablowitz--Ladik equation, it is easier to use
the direct, zero-curvature, approach of the works
\cite{AL3,Suris,Suris2}.

\section{Conservation laws}  \label{sec-cl}

As we mentioned in the Introduction, in this section we discuss an
example of the practical applications of the functional representation of
the ALH and consider once more the question of describing the infinite
series of the conservation laws of this integrable model.

The conserved quantities of the ALH have been known since the work
of Ablowitz and Ladik~\cite{AL2} and one can find there (or in a
textbook; see, e.g., \cite{AS}) how to derive them as well as
their generating function. However, this generating function is
usually given in terms of the solutions of the auxiliary
problem~(\ref{zcr-sp}) (i.e.\ the Jost functions) while one usually wants
to write the conservation laws in terms of the solutions of the
equations of the hierarchy themselves.  Now we know how, using
Miwa's shifts, to express the former in terms of the latter. So we
can describe the conserved quantities (and corresponding
divergence-like formulae) using directly the tau-functions of the
ALH.

The first constants of motion are given by
\begin{gather}
  I_{1} = \sum_{n} \; r_{n-1}q_{n},
\label{integral-1}
\\
  I_{2} = \sum_{n} \;
    r_{n-1}p_{n}q_{n+1} -
    {1 \over 2} \, r_{n-1}^{2} q_{n}^{2}.
\label{integral-2}
\end{gather}
One can easily observe that the summands in these formulae are nothing
but the quantities $\delta_{n}^{j}$ from (\ref{v-j}), {\it videlicet}
\begin{gather}
\delta_{n}^{1} = i r_{n-1}q_{n},
\\
\delta_{n}^{2} =
  i r_{n-2}p_{n-1}q_{n} + i r_{n-1}p_{n}q_{n+1} - i r^{2}_{n-1}q^{2}_{n}.
\end{gather}
Indeed, as follows from (\ref{dq-j}), (\ref{dr-j}) and (\ref{zcr-b-j})--(\ref{zcr-d-j}),
 one can present the $\delta_{n}^{j}$ as
\begin{equation}
\delta_{n}^{j} = \partial_{j} \ln {\tau_{n} \over \tau_{n-1} }.
\end{equation}
Hence
\begin{equation}
  \partial_{k}\delta_{n}^{j} =
  \partial_{jk} \ln \tau_{n} -
  \partial_{jk} \ln \tau_{n-1}.
\end{equation}
Obviously the right-hand side of this identity vanishes after
summation over $n$ in the case of an infinite chain under the zero
or finite-density boundary conditions (or in a periodic case).
Thus the series $d_{n}(\zeta) = \sum\limits_{j=1}^{\infty} \zeta^{j}
\delta_{n}^{j}$ (see (\ref{abcd})) can be viewed as a generating
function for the constants of motion.  To complete the description
of the conservation laws,
\begin{equation}
  \partial_{k}\delta_{n}^{j} =
  f^{jk}_{n} - f^{jk}_{n-1},
\label{cons-law-jk}
\end{equation}
we have to obtain an expression (and a generating function) for
the quantities $f_{n}^{jk} = \partial_{jk} \ln \tau_{n}$. To this
end consider equation (\ref{par-d}). After some algebraic
manipulations with (\ref{zcr-b})--(\ref{zcr-d}) one can rewrite
(\ref{par-d}) in the form which we need:
\begin{equation}
\partial(\eta) d_{n}(\xi) =  f_{n}(\xi,\eta) - f_{n-1}(\xi,\eta),
\label{cons-law}
\end{equation}
where
\begin{gather}
  f_{n}(\xi,\eta) =
  { \xi\eta \over (\xi - \eta)^{2} }
  \left[
    2 d_{n}(\xi) d_{n}(\eta) - i d_{n}(\xi) - i d_{n}(\eta)\right.\nonumber\\
\left.\phantom{f_{n}(\xi,\eta) =} {}  + \xi^{-1}  b_{n}(\xi)  c_{n}(\eta)
    + \eta^{-1} b_{n}(\eta) c_{n}(\xi)
  \right].
\end{gather}
This expression, after expansion in $\xi$ and $\eta$, gives
(\ref{cons-law-jk}) with
\begin{equation}
f^{jk}_{n} =
- \sum_{p=1}^{j}\sum_{q=1}^{k}
\left[
  \delta_{n+1}^{p+q-1}\delta_{n+1}^{j+k+1-p-q} +
  \beta_{n+1}^{p+q}\gamma_{n+1}^{j+k+1-p-q}
\right].
\end{equation}
These formulae solve our problem. However, one may feel some
dissatisfaction with their form. The results obtained can be
written in a more elegant way if one expresses them in the terms
of the tau-functions with shifted arguments. To do this one needs
the following formulae for $b_{n}(\zeta)$, $c_{n}(\zeta)$ and
$d_{n}(\zeta)$ which can be derived from (\ref{zcr-d}) and
(\ref{db-q}), (\ref{dc-r}):
\begin{gather}
b_{n}(\zeta) =
   - d_{*}(\zeta) \;
   { \rho_{n-1}( z - i[\zeta] ) \; \tau_{n}( z + i[\zeta] )
    \over
     \tau_{n-1}( z ) \; \tau_{n}( z ) },
\\
c_{n}(\zeta) =
   - d_{*}(\zeta) \;
   { \tau_{n-1}( z - i[\zeta] ) \; \sigma_{n}( z + i[\zeta] )
    \over
     \tau_{n-1}( z ) \; \tau_{n}( z ) },
\\
d_{n}(\zeta) =
   d_{*}(\zeta)   \;
   { \rho_{n-1}( z - i[\zeta] ) \; \sigma_{n}( z + i[\zeta] )
    \over
     \tau_{n-1}( z ) \; \tau_{n}( z ) }.
\end{gather}
Here $d_{*}(\zeta)$ is some constant (with respect to the index $n$) which
appears when one solves (\ref{zcr-d}) and which depends on the boundary
conditions for the $q_{n}$ and $r_{n}$. Using these relations together
with the functional equations (\ref{tau-fe-t})--(\ref{tau-fe1-r}) and
the `determinant' identities (\ref{two-shift-w})
one can rewrite the conservation laws as follows:
\begin{equation}
\partial(\eta) \mathcal{H}_{n}(z, \bar z; \xi) =
\mathcal{F}_{n}(z, \bar z; \xi, \eta) -
\mathcal{F}_{n-1}(z, \bar z; \xi, \eta),
\label{cons-law-hf}
\end{equation}
where
\begin{equation}
\mathcal{H}_{n}(z, \bar z; \xi) =
{  \rho_{n-1}   ( z - i[\xi], \bar z ) \;
   \sigma_{n} ( z + i[\xi], \bar z )
   \over
   \tau_{n-1}(z, \bar z) \tau_{n}(z, \bar z)
}
\end{equation}
and
\begin{equation}
\mathcal{F}_{n}(z, \bar z; \xi, \eta) =
C(\xi,\eta) \;
{  \rho_{n-1}   ( z - i[\xi] - i[\eta], \bar z ) \;
   \sigma_{n+1} ( z + i[\xi] + i[\eta], \bar z )
   \over
   \tau_{n}^{2}(z, \bar z)
}
\end{equation}
with
\begin{equation}
C(\xi,\eta) =
- d_{*}(\eta)
  \left[ { \kappa(\xi, \eta) \over \xi - \eta } \right]^{2}.
\end{equation}
Thus we have derived the generating function for all conservation laws of
{\it all} equations of the `positive' subhierarchy,
\begin{equation}
  \partial_{k} \mathcal{H}^{j}_{n} =
  \mathcal{F}^{jk}_{n} - \mathcal{F}^{jk}_{n-1},
\label{cons-law-hfjk}
\end{equation}
where
\[
\mathcal{H}_{n}(\xi) = \sum\limits_{j} \xi^{j} \mathcal{H}^{j}_{n},\qquad
\mathcal{F}_{n}(\xi,\eta) =
\sum\limits_{jk} \xi^{j}\eta^{k}\mathcal{F}^{jk}_{n}.
\]
 Note that the
constants of motion,~$I_{i}$, are surely common for all equations of
the ALH, but the form of the corresponding conservation law
(\ref{cons-law-hfjk}) depends upon the flow (operator~$\partial_{k}$) with which we are dealing.

If we restrict ourselves to the zero boundary conditions, $q_{n},
r_{n} \to 0$ as $n \to \pm\infty$, then
$\kappa(\xi,\eta)=\xi-\eta$, $d_{*}(\xi)=i\xi$ and the above
formulae can be simplified to become
\begin{equation}
i\partial_{1} \,
{ \rho_{n-1}^{-} \sigma_{n}^{+} \over \tau_{n} \tau_{n-1} } =
{ \rho_{n-1}^{-} \sigma_{n+1}^{+} \over \tau_{n}^{2} } -
{ \rho_{n-2}^{-} \sigma_{n}^{+} \over \tau_{n-1}^{2} },
\label{cons-law-1}
\end{equation}
where
\begin{equation}
\omega_{n}^{\pm}(z, \bar z) = \omega_{n}( z \pm i[\xi], \bar z).
\end{equation}
Considering (\ref{cons-law-1}) as an equation for power series in
$\xi$ and equating coefficients of $\xi^{j}$, $j=1,2,\ldots$, one can
obtain the infinite series of conservation laws for the discrete
nonlinear Schr\"odinger equation and the discrete modified KdV
equation. The first of them is
\begin{equation}
i\partial_{1} \, r_{n-1}q_{n} =
r_{n-1}p_{n}q_{n+1} - r_{n-2}p_{n-1}q_{n},
\end{equation}
i.e.\ the conservation law corresponding to the constant $I_{1}$
given by (\ref{integral-1}).

Analogous calculations can be carried out for the `negative' subhierarchy.

\section{ALH and other hierarchies}      \label{sec-examples}

It has been shown in \cite{sigma,2dtl,DS} that the ALH possesses some kind
of `universality': many integrable models can be `embedded' into the ALH,
i.e.\ presented as differential consequences of its equations. In this
section we give a few examples of how these interrelations
between the ALH and other integrable systems manifest themselves in the
framework of the functional equations. We derive, starting from the
functional representation of the ALH, functional representation of some
other well-known integrable equations and hierar\-chies.

\subsection{Functional representation of the derivative NLS hierarchy}

The functional equations for the tau-functions
(\ref{tau-fe-s})--(\ref{tau-fe1-t})
can be presented as
\begin{gather}
q_{n}^{+} - q_{n} =
   \zeta \;
   { \tau_{n-1} \over \tau_{n} }
   { \sigma_{n+1}^{+} \over \tau_{n}^{+} },
\label{dnls-1-s}
\\
r_{n-1} - r_{n-1}^{+} =
   \zeta \;
   { \rho_{n-2} \over \tau_{n-1} }
   { \tau_{n}^{+} \over \tau_{n-1}^{+} },
\label{dnls-1-r}
\\
1 + \zeta r_{n-1}q_{n}^{+} =
   { \tau_{n} \over \tau_{n-1} }
   { \tau_{n-1}^{+} \over \tau_{n}^{+} },
\label{dnls-1-t}
\end{gather}
where
\begin{equation}
f_{n}^{\pm} = f_{n}( z \pm i[\zeta] ).
\end{equation}
Shifting the arguments in the first equation,
$z_{k} \to z_{k} - i \zeta^{k}/k $,
and using (\ref{dnls-1-t}) one can rewrite (\ref{dnls-1-s}) and
(\ref{dnls-1-r}) as
\begin{gather}
  \zeta^{-1}
  \left( 1 + \zeta q_{n} r_{n-1}^{-} \right)
  \left( q_{n} - q_{n}^{-} \right) =
  { \tau_{n-1} \sigma_{n+1} \over \tau_{n}^{2} },
\label{dnls-2-q}
\\
  \zeta^{-1}
  \left( 1 + \zeta q_{n}^{+} r_{n-1} \right)
  \left( r_{n-1} - r_{n-1}^{+} \right) =
  { \rho_{n-2} \tau_{n} \over \tau_{n-1}^{2} }.
\label{dnls-2-r}
\end{gather}
It should be noted that the left-hand sides of these two identities, for
fixed~$n$, contain only two functions, $q_{n}$ and $r_{n-1}$, taken at the
different values ot their arguments, while the right-hand sides do not
depend on~$\zeta$. Thus differentiating (\ref{dnls-2-q}) and
(\ref{dnls-2-r}) with respect to~$\zeta$, one can obtain that the
quantities
\begin{equation}
Q=q_{n}
\qquad
\mathrm{and}
\qquad
R=r_{n-1}
\end{equation}
satisfy the system
\begin{gather}
  {\mbox{d} \over \mbox{d}\zeta} \;
  \zeta^{-1}
  \left( 1 + \zeta Q R^{-} \right)
  \left( Q - Q^{-} \right) = 0,
\label{dnls-q}
\\
  {\mbox{d} \over \mbox{d}\zeta} \;
  \zeta^{-1}
  \left( 1 + \zeta Q^{+} R \right)
  \left( R - R^{+} \right) = 0.
\label{dnls-r}
\end{gather}
In such a way we have shown that solutions of the ALH solve also
(\ref{dnls-q}) and (\ref{dnls-r}). This closed system of functional
equations can be viewed as (formal) series in~$\zeta$.
Expanding~(\ref{dnls-q}) and (\ref{dnls-r})  in $\zeta$ and equating
coefficients of different powers of $\zeta$ to zero one can obtain an
infinite number of partial differential equations. The simplest of them are
\begin{gather}
  \partial_{2} Q - i\partial_{11} Q + 2 \, QR \, \partial_{1} Q = 0,
\label{dnlse-q}
\\
  \partial_{2} R + i\partial_{11} R + 2  \, QR  \, \partial_{1} R = 0,
\label{dnlse-r}
\end{gather}
which is the derivative NLS system. Others of these equations are
compatible with (\ref{dnlse-q}) and (\ref{dnlse-r}) (by construction) and can
be transformed to the form
\begin{gather}
\partial_{j} Q =
   \mathcal{P}_{Q} \left( Q, R, \partial Q, \partial R, \dots \right),
\\
\partial_{j} R =
   \mathcal{P}_{R} \left( Q, R, \partial Q, \partial R, \dots \right),
\end{gather}
where $\partial = \partial/\partial z_{1}$, and $\mathcal{P}_{Q}$ and
$\mathcal{P}_{R}$ are polynomials in $Q$, $R$ and their derivatives with
respect to $z_{1}$. Hence one can conclude that they are nothing other
but the higher equations from the hierarchy which begins with
(\ref{dnlse-q}) and (\ref{dnlse-r}), i.e.\ the functional equations
(\ref{dnls-q}) and (\ref{dnls-r}) can be viewed as representing the
derivative NLS hierarchy.

\subsection{Functional representation of the AKNS hierarchy}

To derive the functional equations corresponding to the AKNS hierarchy
consider the `four-point' (or `two-shift') formulae (\ref{two-shift-w}).
After some calculation one can obtain the identities
\begin{equation}
  (\zeta - \xi) \hat Q_{n}^{+} =
  \left( 1 + \zeta\xi \hat Q_{n}^{+} R_{n} \right)
  \left( \zeta Q_{n}^{+} - \xi \hat Q_{n} \right)
\label{akns-1-q}
\end{equation}
and
\begin{equation}
  (\zeta - \xi) \check R_{n}^{-} =
  \left( 1 + \zeta\xi Q_{n} \check R_{n}^{-} \right)
  \left( \zeta R_{n}^{-} - \xi \check R_{n} \right)
\label{akns-1-r}
\end{equation}
for the quantities
\begin{equation}
  Q_{n}={ \sigma_{n+1} \over \tau_{n} }
\qquad
\mathrm{and}
\qquad
  R_{n}={ \rho_{n-1} \over \tau_{n} }.
\label{akns-qr}
\end{equation}
Here
\begin{equation}
  f_{n}^{\pm} = f_{n}( z \pm i[\zeta] ),
\qquad
  \hat f_{n} = f_{n}( z + i[\xi] ),
\qquad
  \check f_{n} = f_{n}( z - i[\xi] ).
\end{equation}
Expansion of (\ref{akns-1-q}) and (\ref{akns-1-r}) in double series in $\zeta$
and $\xi$ leads to an infinite number of partial differential equations.
It can be shown that equations we need (the NLS equation and its higher
analogues) appear as cofactors of the
$\zeta\xi\left(\zeta^{m}-\xi^{m}\right)$-terms, while coefficients of all other terms
can be expressed as their linear combinations. We not discuss further
the double series representation because in our case it can be simplified
and reduced to one similar to (\ref{dnls-q}) and (\ref{dnls-r}). Indeed,
after shifting the arguments of all functions in (\ref{akns-1-q}), $z_{k}
\to z_{k} - i\zeta^{k}/k - i\xi^{k}/k$, one can rewrite it in the $\xi \to
0$ limit as
\begin{equation}
  Q^{-} - \zeta^{2} Q^{2} R^{-} = Q - i\zeta \partial_{1} Q.
\end{equation}
The right-hand side of this equation, which is of no interest for
our purposes and which is linear in $\zeta$, can be eliminated by
differentiating with respect to $\zeta$. This leads to the
following, rather compact relation
\begin{equation}
  {\mbox{d}^{2} \over \mbox{d}\zeta^{2}} \;
  \left( Q^{-} - \zeta^{2} Q^{2} R^{-} \right) = 0.
\label{akns-2-q}
\end{equation}
Analogously, equation (\ref{akns-1-r}) can be transformed to
\begin{equation}
  {\mbox{d}^{2} \over \mbox{d}\zeta^{2}} \;
  \left( R^{+} - \zeta^{2} Q^{+} R^{2} \right) = 0.
\label{akns-2-r}
\end{equation}
Expanding (\ref{akns-2-q}) and (\ref{akns-2-r}) in power series in $\zeta$
one can subsequently obtain the NLS equation,
\begin{gather}
i\partial_{2} Q + \partial_{11} Q + 2 Q^{2} R  = 0,
\\
  -i\partial_{2} R + \partial_{11} R + 2 Q R^{2} = 0,
\end{gather}
the third-order NLS equation,
\begin{gather}
  \partial_{3} Q + \partial_{111} Q + 6 QR \partial_{1}Q = 0,
\\
  \partial_{3} R + \partial_{111} R + 6 QR \partial_{1}R = 0
\end{gather}
and so on, which enables us to identify (\ref{akns-2-q}) and
(\ref{akns-2-r}) with the AKNS hierarchy.

\subsection{Functional representation of the Davey--Stewartson hierarchy}

In this paper (as well as in the paper \cite{1}) we discussed the
'positive' and `negative' subhierarchies separately. At the same time it
is already known that the combined action of both types of the ALH flows,
$\partial / \partial z_{j}$ and $\partial / \partial \bar z_{k}$, leads to
some interesting equations, first of all the Davey--Stewartson system
(see~\cite{DS}) and the 2D Toda lattice (see~\cite{2dtl}). Now we
demonstrate some consequences of the functional equations related to both
subhierarchies considered together.

Equations (\ref{tau-fe-s}) and (\ref{tau-fe-r}),
\begin{gather}
q_{n} - q_{n}^{-}=
  \zeta { \tau_{n-1}^{-} \sigma_{n+1} \over \tau_{n}^{-} \tau_{n} },
\\
r_{n} - r_{n}^{+} =
  \zeta { \rho_{n-1} \tau_{n+1}^{+} \over \tau_{n} \tau_{n}^{+} },
\end{gather}
where the designation
\begin{equation}
f_{n}^{\pm} = f_{n}( z \pm i[\zeta], \bar z )
\end{equation}
is used, can be rewritten, by means of (\ref{tau-fe1-t}), as
\begin{gather}
q_{n}^{-} - \zeta^{2} q_{n} Q_{n} R_{n}^{-} =
  q_{n} - \zeta { \tau_{n-1} \sigma_{n+1} \over \tau_{n}^{2} },
\\
r_{n}^{+} - \zeta^{2} r_{n} Q_{n}^{+} R_{n} =
  r_{n} - \zeta { \rho_{n-1} \tau_{n+1} \over \tau_{n}^{2} }
\end{gather}
with $Q_{n}$ and $R_{n}$ being defined by (\ref{akns-qr}). As in
the examples above the right-hand side of these relations can be
eliminated with the help of the $\mbox{d} / \mbox{d}\zeta$
operator. To obtain a closed system it remains to express the
quantity $Q_{n} R_{n}^{-}$ in terms of $q_{n}$ and $r_{n}$. This
can be done using the `negative' functional equations for the
tau-functions. Indeed, from (\ref{tau-nfe1-s}), (\ref{tau-nfe1-r})
and (\ref{tau-nfe-t}), after some calculations omitted here one
can derive the following relation:
\begin{equation}
- i \bar\partial \left(Q R^{-}\right) =
\zeta^{-1} \left( q r - q^{-} r^{-} \right),
\qquad
\bar\partial = \partial / \partial \bar z_{1}
\end{equation}
or
\begin{equation}
Q R^{-} =
i \zeta^{-1} \bar\partial^{-1} \left( q r - q^{-} r^{-} \right)
\end{equation}
which leads to
\begin{equation}
{\mbox{d}^{2} \over \mbox{d}\zeta^{2}} \; \left[    q^{-} +
   i \zeta \, q \, \bar\partial^{-1} \left( q^{-} r^{-} - q r \right)
\right] = 0
\label{ds-2-q}
\end{equation}
and
\begin{equation}
{\mbox{d}^{2} \over \mbox{d}\zeta^{2}} \;
  \left[  r^{+} +
    i \zeta \, r \, \bar\partial^{-1} \left( q r - q^{+} r^{+} \right)
  \right] = 0.
\label{ds-2-r}
\end{equation}
The lowest-order partial differential equations corresponding to this
system, i.e.\ the terms proportional to $\zeta^{2}$ in square brackets, can
be presented as
\begin{gather}
i\partial_{2} q + \partial_{11} q - 2 C q  = 0,
\\
  -i\partial_{2} r + \partial_{11} r - 2 C r = 0,
\\
  \bar\partial C = \partial q r.
\end{gather}
It is not difficult to show that these equations, which describe
an integrable system proposed by Zakharov~\cite{Z}, together with
their `negative' counterparts are equivalent to the
Davey--Stewartson system (see~\cite{DS}). Thus one can consider
(\ref{ds-2-q}) and (\ref{ds-2-r}) as functional equations
representing the Davey--Stewartson hierarchy. Similar
representation has been derived earlier by Bogdanov and
Konopelchenko using the analytic-bilinear approach~\cite{BK}.

\section{Conclusion}

The ALH is usually considered as a set of differential-difference
equations. However, the differential `component' of the hierarchy
can be viewed as a consequence of the difference one. It is
clearly seen in the framework of the inverse scattering transform.
If one fixes the $U$-matrix in (\ref{zcr-sp}) and restricts
oneself to the polynomial $V$-matrices, one almost completely
determines the latter, i.e.\ all differential equations of the
hierarchy are 'hidden' in the difference problem (\ref{zcr-sp}).
Thus it is possible to obtain a wide range of results related to
the hierarchy without dealing with separate flows. One of the main
aims of this paper was to embody this idea (Kyoto approach~\cite{JM}).
The functional equations are a useful tool for such
`global' consideration: an infinite set of differential equations
(continuous flows) is replaced with one equation (discrete flow).
Another motive behind this work was the following one. The ALH has
a rather long history. Much effort has been spent to study the
discrete NLS and discrete modified KdV equations, the simplest
equations of the ALH. However, the general, abstract, theory of
this hierarchy (say, similar to the well-elaborated KP theory) has
not yet been developed. The results presented above seem to give a
perspective from this viewpoint. For example the representation~(\ref{pi-plus})
can be a starting point to investigate the
group-algebraic structures behind the ALH. The appearance of
Miwa's shifts in the main equations of this paper, together with
the Plu\"cker-like relations~(\ref{det-w}), indicates that we are
close to the description of the ALH using the language of the
Grassmannians and it is interesting to continue the studies in
this direction, to introduce vertex operators, to construct the
free-field realization of the ALH and to address other questions
which are of much interest in the modern theory of the integrable
systems.

\subsection*{Acknowledgements}

This work was partly carried out during the author's stay at the Abdus
Salam International Centre for Theoretical Physics which is
gratefully acknowledged for its kind hospitality and has been
supported by the Ministerio de Ciencia, Ciltura y Deporte of Spain
under grant SB1999-AH777133. I am also grateful to the referee of
this paper for careful reading the manuscript and making a few
crucial remarks.

\label{Vekslerchik-lastpage}

\end{document}